\documentclass[a4paper]{an}

\usepackage{times}
\usepackage{amssymb}
\usepackage{graphicx}
\begin{document}
\title{Element spots in Ap and Hg-Mn stars from current-driven diffusion} 
\author{V.~Urpin\inst{1,2}}
\institute{$^{1)}$ INAF, Osservatorio Astrofisico di Catania,
           Via S.Sofia 78, 95123 Catania, Italy \\
           $^{2)}$ A.F.Ioffe Institute of Physics and Technology,
           194021 St. Petersburg, Russia
}

\date{\today}
\abstract
{The stars of the middle main sequence often have spot-like chemical
structures at their surfaces. 
We consider the diffusion process caused by electric currents 
that can lead to the formation of such chemical spots.  
Diffusion was considered using partial momentum equations 
derived by the Chapman-Enskog method. 
We argue that diffusion caused by electric currents can substantially
change the surface chemistry of stars and form spotted chemical
structures even in a relatively weak magnetic field. The considered
mechanism can be responsible for a formation of element spots in
Hg-Mn and Ap-stars.  
}

\authorrunning{V. Urpin}
\titlerunning{Element spots and current-driven diffusion}

\maketitle

\section{Introduction}

Diffusion can lead to evolution of atmospheric chemistry in stars 
and be the reason of chemical peculiarities. This particularly 
concerns the stars of the middle main sequence that often have 
relatively quiescent surface layers. Many stars with peculiar 
chemical abundances show line variations caused by element spots 
on their surface (see, e.g., Pyper 1969, Khokhlova 1985, 
Silvester et al. 2012). It was thought that chemical spots can 
only occur in the presence of a strong magnetic field. Indeed, 
some Ap stars show variations of both spectral lines and magnetic 
field strength that can be related to rotation of chemical and 
magnetic spots. Often such stars have the strongest concentration 
of heavy elements around the magnetic poles (see, e.g., Havnes 
1975). Note that a reconstruction of the stellar magnetic geometry 
from observations is a very complex problem. The magnetic Doppler 
imaging code developed by Piskunov \& Kochukhov (2002) makes it 
possible to derive the magnetic map of a star self-consistently 
with the distribution of the chemical elements. The reconstructions 
show that the magnetic and chemical maps of stars can be very 
complex (Kochukhov et al. 2004a) and usually chemical elements do 
not exhibit a correlation with the magnetic geometry. The 
calculated distributions demonstrate the complexity of diffusion 
in Ap-stars and show that chemical distributions are affected by 
a number of poorly understood phenomena and are not directly 
related to the strength of the magnetic field.  

Often, the chemical spots on the surface of stars are related to 
anisotropic diffusion in the magnetic field. Indeed, the magnetic 
field of Ap-stars ($B \sim 10^3 - 10^4$ G) is sufficiently strong 
to magnetize plasma and make diffusion anisotropic. Anisotropy of 
diffusion is characterized by the Hall parameter, $x = \omega_{Be} 
\tau_e$, where $\omega_{Be} = e B/m_e c$ is the gyrofrequency of 
electrons and $\tau_e $ is their relaxation time. If the 
background plasma is hydrogen, then $\tau_e = 3 \sqrt{m_e} 
(k_b T)^{3/2}/4 \sqrt{2 \pi} e^4 n \Lambda $ (see, e.g., Spitzer 
1998) where $n$ and $T$ are the number density of electrons and 
their temperature, $\Lambda$ is the Coulomb logarithm. If $x 
\geq 1$, the rates of diffusion along and across the magnetic 
field are different and diffusion can lead to inhomogeneous 
element distributions. The condition $x \geq 1$ yields the 
following estimate of the magnetic field that magnetizes plasma
\begin{equation}
B \geq B_e = 2.1 \times 10^3 n_{15} T_4^{-3/2} \Lambda_{10} \;\; 
{\rm G}, 
\end{equation}      
where $\Lambda_{10} = \Lambda/10$, $n_{15} = n/10^{15}$, and $T_{4} 
= T/10^4 K$. Some stars with chemical spots have such a strong 
magnetic field and diffusion can be anisotropic there.     

In recent years, however, the discovery of chemical inhomogeneities
in the so-called Hg-Mn stars has rised some doubts regarding their
magnetic origin. The aspect of spot-like chemical structures in 
HgMn stars was discussed first by Hubrig \& Mathys (1995). 
In contrast to Ap-stars, no strong madnetic field of kG order has 
ever been detected in HgMn stars. For instance, Wade et al. (2004) 
find no longitudinal field above 50 G in the brightest Hg-Mn star 
$\alpha$And with chemical spots at the surface. The authors 
establish an upper limit of the global field at $\approx 300$ G 
that is not sufficient to magnetize plasma. Weak magnetic fields 
in the atmospheres of Hg-Mn stars have been detected by a number 
of authors (see, e.g., Hubrig \& Castelli 2001, Hubrig et al. 
2006, Makaganiuk et al. 2011, 2012). In a recent study by Hubrig 
et al. (2012), the previous measuments of the magnetic field 
have been re-analysed and the presence of a weak longitudinal 
magnetic field up to 60-80 G have been revealed in several HgMn 
stars. The complex interrelations between the magnetic field and 
the chemical structures show how incomplete is our understanding 
of diffusion processes in stars.

In this paper, we consider one more diffusion mechanism 
that contributes to a formation of chemical spots in stars.
This mechanism is relevant to electric currents and has not 
been studied in stellar conditions yet. We concentrate on the 
main qualitative features of this process and compare the
diffusion rate caused by the presence of electric currents and
the rate of other diffusion processes. We show that interaction 
of the electric current with ions leads to diffusion in the 
direction perpendicular to the both electric current and 
magnetic field. Such diffusion can alter the surface chemical 
distributions at a substantially weaker magnetic field than $B_e$. 

\section{Basic equations}

Consider a cylindrical plasma configuration with the magnetic
field parallel to the axis $z$, $\vec{B} = B \vec{e}_{z}$; $(s, 
\varphi, z)$ and $(\vec{e}_s, \vec{e}_{\varphi}, \vec{e}_{z})$ are 
cylindrical coordinates and the corresponding unit vectors. We 
assume that the magnetic field depends on the cylindrical radius
alone, $B=B(s)$. Then, the electric current is 
\begin{equation}
j_{\varphi} = - (c/4 \pi) \partial B/\partial s.  
\end{equation}
We suppose that $j_{\varphi} \rightarrow 0$ at large $s$ and,
hence, $B \rightarrow B_0$=const at $s \rightarrow \infty$.
Note that $B(s)$ can not be an arbitrary function of $s$ because, 
generally, the magnetic configurations can be unstable for
some dependences of $B(s)$ on $s$ (see, e.g., Tayler 1973, 
Bonanno \& Urpin 2008a,b for more detail). The timescale of 
this instability is usually shorter than the diffision 
timescale and, therefore, a formation of chemical structures 
in such magnetic configurations is unlikely. 

We assume that plasma consists of electrons $e$, protons
$p$, and a small admixture of heavy ions $i$. The number density 
of species $i$ is such small that it does not influence dynamics 
of plasma. Therefore, this species can be treated as test 
particles that interacts only with electrons and background 
hydrogen plasma. The hydrostatic equilibrium reads
\begin{equation}
- \nabla p + \rho \vec{g} + \frac{1}{c} \vec{j} \times 
\vec{B} = 0,
\end{equation}
where $p$ and $\rho$ are the pressure and density, respectively, 
$\vec{g} = -g \vec{e}_z$ is gravity. Since the background 
plasma is hydrogen, $p \approx 2n k_B T$ where
$k_b$ is the Boltzmann constant. Integrating the $s$-component 
of Eq.~(3), we obtain 
\begin{equation}
n = n_0 \left( \frac{T_0}{T} \right) \left( 1 + \frac{1}{\beta_0}
- \frac{1}{\beta} \right),
\end{equation}
where $\beta = 8 \pi p_0 / B^2$; $(p_0, n_0, T_0, \beta_0)$ are 
the values of $(p, n, T, \beta)$ at $s \rightarrow \infty$.

The partial momentum equations in fully ionized multicomponent 
plasma has been considered by a number of authors (see, e.g.,
Urpin (1981)). If the mean hydrodynamic velocity is zero and 
only small diffusive velocities are non-vanishing, the partial 
momentum equation for the species $i$ can be written as
\begin{equation}
- \nabla p_i + Z_i e n_i \left( \vec{E} + 
\frac{\vec{V}_i}{c} 
\times \vec{B} \right) + \vec{R}_{i e} + \vec{R}_{i H} + 
\vec{F}_i
= 0, 
\end{equation}
where $Z_i$ is the charge number of the species $i$, $p_i$, and 
$n_i$ are its partial pressure and number density, $\vec{V}_i$ 
is its velocity, and $\vec{E}$ is the electric field. The force 
$\vec{F}_i$ is the external force on species $i$; in stars, 
$\vec{F}_i$ is usually determined by gravity $\vec{g}$ and the 
radiation force. The forces $\vec{R}_{i e}$ and $\vec{R}_{i H}$ 
are caused by the interaction of ions $i$ with electrons 
and protons, respectively. Note that forces $\vec{R}_{i e}$ 
and $\vec{R}_{i H}$ are internal and their sum over all plasma 
components is zero in accordance with Newton's third law. 
Since diffusive velocities are typically small, we neglect 
the terms proportional $(\vec{V}_i \cdot \nabla) \vec{V}_1$ in 
the momentum equation (5).

The $s$- and $\varphi$-components of Eq.(5) yield
\begin{eqnarray}
- \frac{d}{ds} (n_i k_B T) + Z_i e n_i \left( \! E_s \! + \!
\frac{V_{i \varphi}}{c} B \! \right) + R_{ie s} + R_{iH s} 
= 0, \\
Z_i e n_i \left( \! E_{\varphi} \! - \! \frac{V_{i s}}{c}  B \! \right) + 
R_{ie \varphi} + R_{iH \varphi} = 0.
\end{eqnarray}
The force $\vec{R}_{ie}$ is caused by scattering of ions $i$ on 
electrons. If $n_i$ is small compared to the number density of 
protons, $\vec{R}_{ie}$ is given by      
\begin{equation}
\vec{R}_{ie} = - \frac{Z_i^2 n_i}{n} \vec{R}_{e}
\end{equation}
where $\vec{R}_{e}$ is the force acting on the electron
gas (see, e.g., Urpin 1981). Since $n_i \ll n$, 
$\vec{R}_{e}$ is determined mainly by scattering of electrons 
on protons but scattering on ions $i$ gives a small 
contribution. Therefore, we can use for $\vec{R}_{e}$ the 
expression for one component hydrogen plasma calculated by 
Braginskii (1965). In our model of a cylindrical
plasma configuration, this expression reads
\begin{equation}
\vec{R}_{e} = - \alpha_{\perp} \vec{u} + \alpha_{\wedge} \vec{b}
\times \vec{u} - \beta^{uT}_{\perp} \nabla T - \beta^{uT}_{\wedge}
\vec{b} \times \nabla T,
\end{equation}   
where $\vec{u} = - \vec{j}/en$ is the difference between the mean 
velocities of electrons and protons,  $\vec{b} = \vec{B}/B$, 
$\alpha_{\perp}$, $\alpha_{\wedge}$, $\beta^{uT}_{\perp}$, and 
$\beta^{uT}_{\wedge}$ are coefficients calculated by Braginskii 
(1965). The first two terms on the r.h.s. of Eq.(9) describe the
standard friction force caused by a relative motion of the electron 
and proton gases. The last two terms on the r.h.s. of Eq.(9)
represent the so-called thermoforce caused by a temperature 
gradient. This part of $\vec{R}_{e}$ is responsible for 
thermodiffusion. 

Taking into account Eq.(2), we have 
\begin{equation}
\vec{u} = \frac{c}{4 \pi e n} \frac{d B}{d s} \vec{e}_{\varphi}.
\end{equation}
In this paper, we consider diffusion only in a relatively weak 
magnetic field that does not magnetize electrons, $x \ll 1$. 
Substituting Eq.(10) into Eq.(8) and using coefficients 
$\alpha_{\perp}$, $\alpha_{\wedge}$, $\beta^{uT}_{\perp}$, and 
$\beta^{uT}_{\wedge}$ with the accuracy in linear terms in $x$, 
we obtain 
\begin{eqnarray}
R_{ie \varphi} = Z_i^2 n_i \left( 0.51 \frac{m_e}{\tau_e} u
+ 0.81 x k_B  \frac{d T}{d s} \right), \\
R_{ie s} = Z_i^2 n_i \left( 0.21 x \frac{m_e}{\tau_e} u
+ 0.71 k_B \frac{d T}{d s} \right). 
\end{eqnarray}

The force $\vec{R}_{iH}$ is the sum of two terms as well, $\vec{R}_{iH}
=\vec{R}^{'}_{iH} + \vec{R}^{''}_{iH}$ that are proportional to the 
relative velocity of ions $i$ and protons and the temperature gradient,
respectively. The friction force $\vec{R}^{'}_{iH}$ can be easily
calculated if $A_i = m_i/m_p \gg 1$. In this case, $\vec{R}^{'}_{iH} 
\propto (\vec{V}_H - \vec{V}_i)$ but taking into account that the 
mean velocity of the background plasma is zero in our model, the 
friction force can be represented as (see, e.g., Urpin 1981) 
\begin{equation}
\vec{R}^{'}_{iH} \approx \frac{0.4 2 m_i n_i Z^2_i}{\tau_i} (-\vec{V}_i),
\end{equation} 
where $\tau_i = 3\sqrt{m_i} (k_B T)^{3/2} / 4 \sqrt{2 \pi} e^4 n 
\Lambda$ and $\tau_{i}/ Z^2_i$ is the timescale of ion-proton 
scattering; we assume that Coulomb logarithms are the same for all 
types of scattering.    

The thermal part of the friction force, $\vec{R}^{''}_{iH}$,
has been calculated by Urpin (1981). Since there 
is no diffusion in the $z$-direction, the expression for 
$\vec{R}^{''}_{iH}$ with accuracy in linear terms in magnetization 
can be written as
\begin{equation}
\vec{R}^{''}_{iH} = Z^2_i n_i k_B ( 1.71 \nabla T - 
3.43 y \vec{b} \times \nabla T), 
\end{equation}
where
\begin{equation}
y = \frac{e B \tau_p}{m_p c}, \;\;\;
\tau_p = \frac{3 \sqrt{m_p} (k_B T)^{3/2}}{4 \sqrt{2 \pi} 
e^4 n \Lambda}.
\end{equation}
Then, the cylindrical components of $\vec{R}^{''}_{iH}$ are
\begin{eqnarray}
R_{iH s} = Z^2_i n_i \left( - \frac{m_i}{\tau_i} V_{i s} +
1.71 k_B \frac{d T}{d s} \right), \\
R_{iH \varphi} = Z^2_i n_i \left( - \frac{m_i}{\tau_i} V_{i \varphi} -
3.43 y k_B \frac{d T}{d s} \right).
\end{eqnarray}

The momentun equation for the species $i$ (Eq.(5)) depends
on cylindrical components of the electric field, $E_s$ and $E_{\varphi}$.
These components can be determined from the momentum equations
of electrons and protons
\begin{eqnarray}
- \nabla (n k_B T) - e n \left( \vec{E} + \frac{\vec{u}}{c} \times
\vec{B} \right) + \vec{R}_e = 0, \\
- \nabla (n k_B T) + e n \vec{E} - \vec{R}_e + \vec{F}_p = 0.  
\end{eqnarray}
Taking into account the condition (3) and the friction force 
$\vec{R}_e$  (Eq.~(9)) calculated by Braginskii (1965), we obtain 
with accuracy in linear terms in $x$ 
\begin{eqnarray}
E_s = - \frac{uB}{2c} - \frac{1}{e} \left( 0.21 \frac{m_e u}{\tau_e} x
+ 0.71 k_B \frac{d T}{d s} \right) ,     \\
E_{\varphi} = - \frac{1}{e} \left( 0.51 \frac{m_e u}{\tau_e}  +
0.81 x k_B \frac{d T}{d s} \right) .
\end{eqnarray}

Substituting Eqs.(11)-(12), (16)-(17), and (20)-(21) into Eqs.(6) 
and (7), we arrive at the expression for a diffusion velocity, 
$\vec{V}_i$. The radial component of this velocity reads
\begin{equation}
V_{is} = V_{n_i} + V_T + V_B,
\end{equation}
where
\begin{equation}
V_{n_i} = - D \frac{d \ln n_i}{d s}, \;\; V_T = D_T \frac{d \ln T}{d s},
\;\; V_B = D_B \frac{d \ln B}{d s};
\end{equation} 
$V_{ni}$ and $V_T$ are the velocities of ordinary diffusion and 
thermodiffusion, respectively, $V_B$ is the diffusive velocity 
associated with the electric current. The diffusion coefficients  
are
\begin{eqnarray}
D = \frac{2.4 c_i^2 \tau_i}{Z_i^2}, \;\;\; D_T = \frac{2,4 c_i^2 \tau_i}{Z_i^2}
(2.42 Z_i^2 - 0.71 Z_i -1 ), \nonumber \\
D_B = \frac{2.4 c_A^2 \tau_i}{A_i Z_i} (0.21 Z_i - 0.71), 
\end{eqnarray} 
where $c_i^2 = k_B T/m_i$ and $c_A^2 = B^2 / 4 \pi m_p n$. Eq.(22) 
describes the drift of ions $i$ under the combined influence of 
$\nabla n_i$, $\nabla T$, and $\vec{j}$. 

The diffusive velocity given by Eq.~(22) differs from the standard
expression used in astrophysical calculations (see, e.g., Chapman \&
Cowling 1970, Burgers 1969) by the presence of a term $\propto dB/ds$.
It follows from our consideration that this term is caused by scattering
of heavy ions on electrons. The classical works by Chapman \& Cowling
(1970) and Burgers (1969) derive the atomic diffusion coefficients from 
the Boltzmann equation but these coefficients are better suited to 
diffusion in neutral gases or plasma with a large charge of the 
background ions. The point is that these studies neglect scattering of 
impurities on electrons in plasma, and take into account their 
scattering only on the background ions. The latter is correct if the 
charge of background ions, $Z_0$, is large, $Z_0 \gg 1$. In stellar 
plasmas, however, the main background ions are usually protons and, 
hence, $Z_0 =1$. Therefore, neglecting the contribution of electrons 
into kinetic processes is unjustified. This fact was first clearly 
understood by Braginskii (1965) in his theory of transport phenomena 
in a high-temperature plasma. This result can be clarified by simple 
qualitative estimates. Indeed, the momentum of electrons is $\sim 
m_e c_e$ ($c_e= \sqrt{k_B T/ m_e}$ is the thermal velocity of electrons), 
and the rate of momentum transfer by electrons to impurities is  $\sim 
m_e c_e \nu_e \sim m_e c_e / \tau_e$ where $\nu_e$ is the frequency of 
electron collisions. On the other hand, the momentum of protons is 
$\sim m_p c_p$ where $c_p= \sqrt{k_B T/ m_p}$ is the thermal velocity 
of protons and, correspondingly, the rate of momentum transfer by 
protons is  $\sim m_p c_p \nu_p \sim m_p c_p / \tau_p$ where $\nu_p$ 
is the frequency of proton collisions. Comparing these expressions, 
we obtain that the rates of momentum transfer by electrons and protons 
are of the same order and, hence, neglecting the electron contribution 
is unjustified in plasma with $Z_0 \sim 1$. However, if the background
plasma consists of ions with the charge $Z_0 \gg 1$, then one should
replace $\tau_p$ by the relaxation time of the background ions that is 
$\sim \tau_p / Z_0^4$. In this case, the rate of momentum transfer by 
ions turns out to be $\sim Z_0^4$ times greater than that by electrons. 
If $Z_0 \gg 1$, the electron contribution is small and can be neglected.
Therefore, the classical diffusion theory is justified in this case.    
          
The fact that the consistent consideration of scattering on electrons
leads to diffusion of heavy ions with the velocity $\propto dB/ds$
is well known in plasma physics and was first discussed by Vekshtein
et al. (1975). This process plays an important role in diffusion of 
impurities from the walls of the discharge chamber and diaphragms in a 
dense plasma in tokamaks (see, e.g., Vekshtein 1987 for review). Even 
a small fraction of impurity ions can considerably affect the radiation, 
electrical conductivity, and other plasma parameter. Unfortunaly, this 
effect is usually neglected in studies of diffusion in stars but we 
will show that it can play an important role in a spot formation,
particularly, in weakly magnetized stars.

\section{Distribution of ions in the presence of electric currents}

Consider the equilibrium distribution of elements in our 
model. In equilibrium, we have $V_{i s} = 0$ and Eq.(22) yields 
\begin{equation}
\frac{d \ln n_i}{d s} = \frac{D_T}{D} \frac{d \ln T}{d s} +
\frac{D_B}{D} \frac{d \ln B}{d s}. 
\end{equation}
The second term on the r.h.s. is caused by the presence of electric 
currents and describes the current-driven diffusion. 
Note that this type of diffusion is driven by 
the electric current rather than an inhomogeneity of the 
magnetic field. 
Ocasionally, the conditions $d B/ ds \neq 0$ 
and $j \neq 0$ are equivalent in our simplified model.
Equation of hydrostatic equilibrium (3) yields
\begin{equation}
\frac{d}{ds}(n k_B T) = - \frac{B}{8 \pi} \frac{d B}{d s}. 
\end{equation}  
Substituting expression (26) into Eq.(25) and integrating, 
we obtain
\begin{equation}
\frac{n_i}{n_{i0}} = \left( \frac{T}{T_0} \right)^{\nu}
\left( \frac{n}{n_0} \right)^{\mu},
\end{equation}
where
\begin{eqnarray}
\nu = 2 Z_i^2 + 0.71 Z_i -1 , 
\nonumber \\
\mu =  - 2 Z_i (0.21 Z_i - 0.71),
\end{eqnarray}
where $n_{i0}$ is the value of $n_i$ at $s \rightarrow \infty$. 
Denoting the local abundance of the element $i$ as $\gamma_i = 
n_i/n$ and taking into account Eq.~(4), we have 
\begin{equation}
\frac{\gamma_i}{\gamma_{i0}} \!=\! \left( \frac{T}{T_0} \right)^{\nu}
\!\!\! \left( \frac{n}{n_0} \right)^{\mu-1} \!\!\! = \!
\left( \frac{T}{T_0} \right)^{\nu -\mu +1} \!\!\!
\left( 1 + \frac{1}{\beta_0} - \frac{1}{\beta} \right)^{\mu -1} \!\!\!\!,
\end{equation}
where $\gamma_{i0} = n_{i0}/n_0$. It turns out that the local 
abundance of ions is determined by both the temperature and 
magnetic field. The dependence of $\gamma_i$ on $T$ is 
very sensitive to the charge number of ions. For example, if 
$Z_i = 1$, the exponent in Eq.(29) is $\nu - \mu +1 = 1.71$ 
but it is as large as 8.26 if $Z_i =2$. Therefore, even a 
small change in the temperature can be the reason of a
significant variation in the local abundance of chemical 
elements. If the magnetic field is constant then abundance 
anomalies are determined by the thermodiffusion alone. In 
this case, we have 
\begin{equation}
\frac{\gamma_i}{\gamma_{i0}}  = 
\left( \frac{T}{T_0} \right)^{2.42 Z_i^{2} - 0.71 Z_i }.
\end{equation}
Therefore, the regions with a higher temperature, $T > T_0$,
should be overabundant by heavy elements but the regions 
with a lower temperature should be underabundant. 

Local abundances are also flexible to the field strength 
and, particularly, this concerns very heavy ions. If 
variations of the temperature are neglidgible and 
$T \approx T_0$, then the distribution of elements is
determined by the current-driven diffusion alone. In this 
case,
\begin{equation}
\frac{\gamma_i}{\gamma_{i0}} = \left( 1 + \frac{1}{\beta_0} 
- \frac{1}{\beta} \right)^{\mu -1}.
\end{equation}
Note that the exponent $(\mu -1)$ is always negative if $Z_i 
\geq 3$ and, hence, heavy elements with $Z_i \geq 3$ are in 
deficit ($\gamma_i < \gamma_{i0}$) in the region with a weak 
magnetic field ($B > B_0$)
but, on the contrary, such elements should be overabundant
in the spot where the magnetic field is weaker than the
external field $B_0$. The quantity $(\mu -1)$ can reach  
large negative values and, therefore, dependence (31) on the 
magnetic field is very sharp. A combined influence of both 
thermo- and current-driven diffusion can result in a rather 
complicated distribution of elements.

\section{Conclusion}

We have considered diffusion of elements in the surface layers
of stars under a combined influence of different diffusion 
mechanisms. A special attention was paid to the current-driven 
diffusion that has not been discussed yet
in the context of chemical spots on stars.
The diffusion velocity caused by electric current can be
comparable or higher than the velocity of thermodiffusion.
For instance, if electrons are not magnetized ($x < 1$) 
the velocities of thermo- and current-driven diffusions 
can be estimated as 
\begin{equation}
V_T \sim \frac{5 c_i^2 \tau_i}{L_T} , \;\;\;
V_B \sim \frac{c_A^2 \tau_i}{A_i L_B},
\end{equation}
where $L_T$ and $L_B$ are the lengthscales of $T$ and $B$. 
The condition $V_B > V_T$ is satisfied if $c_A \gg 2 c_i 
A_i^{1/2} (L_B/L_T)^{1/2}$ or 
\begin{equation}
B > 2.6 \times 10^2 n_{15}^{1/2} T_{4}^{1/2}
\left( \frac{L_B}{L_T} \right)^{1/2} \;{\rm G},
\end{equation}
where $n_{15} = n/10^{15}$cm$^{-3}$ and $T_4= T/10^4$K. 
It appears that even a relatively weak 
magnetic field ($\sim 10-100$ G) can be the reason of
current-driven diffusion with the velocity greater than that 
of thermodiffusion. From Eq.~(32), one can estimate the
velocity of current-driven diffusion as
\begin{equation}
V_B \sim 1.1 \times 10^{-4} A^{-1/2}_i 
B_4^2 n_{15}^{-2} T_4^{3/2} \Lambda_{10} L_{B \; 10}^{-1} \;\;
{\rm cm/s},
\end{equation}  
where $\Lambda_{10} = \Lambda/10$, $B_4 =B /10^4$ G, and
$L_{B \; 10} = L_B /10^{10}$cm. The velocity $V_B$ turns
out to be sensitive to the field ($\propto B^2$)
and, therefore, diffusion in a weak magnetic field requires 
a longer time to reach equilibrium. 

The considered mechanism can form chemical spots even if 
the magnetic field is relatively weak whereas other diffusion 
processes produce spots only if the magnetic field is 
substantially stronger. For example, the radiative force and 
gravity can generally be responsible for chemical inhomogeneities 
in stars (see, e.g., Vauclair et al. 1979, Michaud et al. 1981). 
The radial diffusion velocity driven by these forces can be 
relatively large, and the distribution of impurities reaches a 
radial equilibrium on a short time scale (Michaud 1970). If the 
radiative and gravitational forces are of the same order of 
magnitude then the velocity of radial diffusion can be estimated as  
\begin{equation}    
V_r \sim g \tau_i
\end{equation}
(see Vauclair et al. 1979). This velocity is typically greater than 
$V_B$ in the surface layers of stars but the radial diffusion cannot 
form chemical spots if the radiative force and $\vec{g}$ have 
spherical symmetry. Departures from sphericity can be caused by the 
magnetic field since the diffusion velocity depends on its direction 
and strength. For instance, the radial diffusion velocities differ by 
a term of the order of 
\begin{equation}
\Delta V \sim V_r (\omega_{Bi} \tau_i /Z_i^2)^2
\end{equation} 
if the magnetic field is parallel and perpendicular to gravity;
$\omega_{Bi} =e B /m_i c$ is the gyrofrequency of impurities (see, 
e.g., Vauclair et al. 1979, Alecian \& Stift 2006). This difference in 
the radial velocities rather than $V_r$ itself leads to formation of 
a spotted structure because spots cannot be formed if $\Delta V =0$.
Usually, $\Delta V$ is much smaller than $V_r$ for more or less 
realistic stellar magnetic fields. For example, using calculations 
of Vauclair et al. (1979), one can estimate that $\Delta V$ is 
comparable to $V_r$ if $B \sim 3 \cdot 10^4$ and $ \sim 10^5$ G at 
the optical depth $1$ and $10$, respectively. These fields are even 
stronger than those detected in Ap-stars. In the case of Hg-Mn stars, 
the magnetic field is likely as weak as 10-100 G and, hence, $\Delta V$ 
is typically $\sim 10^8-10^6$ times smaller than $V_r$. 
Since $\Delta V$ turns out to be small, the velocity of current-driven
diffusion can play an important role in real stars. The velocity 
$V_B$ exceeds $\Delta V$ if the electric current satisfies the
inequality
\begin{equation}
j 
/ \left(c B /4 \pi H \right)  > A_i \frac{c_s^2}{c_A^2} (\omega_{Bi} \tau_i /
Z_i^2 )^2.   
\end{equation}
The parameter $\omega_i \tau_i$ is small in stars and becomes 
greater than 1 if   
\begin{equation}
B > 10^5 n_{15} T_4^{-3/2} \Lambda_{10} \;\; {\rm G}.
\end{equation}
Therefore, the current-driven diffusion can dominate the radiative
diffusion even at a relatively small current.

The current-driven mechanism leads to a drift of ions in
the direction perpendicular to both the magnetic field and
electric current. Therefore, a distribution of chemical 
elements in stars depends essentially on the geometry of 
fields and currents. In the regions where tangential to 
the surface components of the both magnetic field and current 
are greater than normal ones, the considered mechanism 
may lead to the vertical drift of heavy ions. As a result, 
surface layers can be overabundant (or underabundant) by 
heavy element. In the regions where the field is approximately
perpendicular to the surface but the current is tangential
or the current is normal but the field is tangential, heavy
ions drift basically in the tangential direction and can 
form chemical spots.

The mechanism considered can operate in various astrophysical 
bodies where the electric currents are non-vanishing. Like 
other diffusion processes, the current-driven diffusion
can lead to a formation of chemical spots if the star has  
relatively quiescent surface layers. That is the case, for
example, for white dwarfs and neutron stars. Many neutron 
stars have strong magnetic fields and, most likely, topology
of these fields is very complex with spot-like structures at
the surface (see, e.g., Bonanno et al.2005, 2006). As it was
discussed in this paper, such magnetic structures can be
responsible for the formation of a spot-like element distribution 
at the surface. Such chemical structures can be important, for
instance, for the emission spectra, diffusive 
nuclear burning (Chang \& Bildsten 2004, Brown et al. 2002), 
etc. Evolution of neutron stars is very complicated, 
particularly, in binary systems (see, e.g., Urpin et al.
1998) and, as a result, a surface chemistry can be complicated
as well. Diffusion processes may play an important role 
in this chemistry (see, e.g., Brown et al. 2002, Medin \&
Cumming 2014) and can be the reason of chemical sports on 
the surface of these stars.   

\vspace{0.3cm}

\noindent
{\it Acknowledgements}. The author thanks the Russian
Academy of Sciences for financial support under the 
programme OFN-15.

\end{document}